\documentclass{iau} 
\usepackage{graphicx}

\usepackage[english]{babel}
\usepackage[utf8x]{inputenc}
\usepackage{gensymb}
\usepackage{pifont}%
\newcommand{\cmark}{\ding{51}}%
\newcommand{\xmark}{\ding{55}}%

\usepackage{natbib}
\usepackage{color}
 
\newcommand{\sol}[1]{$_\odot$}
\newcommand{\apj}[1]{\textit{ApJ},}
\newcommand{\aap}[1]{\textit{A\&A},}
\newcommand{\apjl}[1]{\textit{ApJ} (Letters),}

\title[Chemical complexity in Serpens] 
{Chemical and kinematic complexity of the very young star-forming region Serpens Main observed with ALMA}

\author[Tychoniec et al.]  
{Łukasz Tychoniec$^1$,
 Charles L. H. Hull$^2$,\\
 John J. Tobin$^3$,
 \and 
  Ewine F. van Dishoeck$^{1,4}$
 }

\affiliation{$^1$Leiden Observatory, Leiden University, PO Box 9513, 2300RA, Leiden, The Netherlands\\ email: {\tt tychoniec@strw.leidenuniv.nl} \\[\affilskip]
$^2$ Harvard-Smithsonian Center for Astrophysics, 60 Garden St.,Cambridge, MA 02138, USA \\[\affilskip]
$^3$Homer L. Dodge Department of Physics and Astronomy, University of Oklahoma, 440 W. Brooks Street, Norman, OK 73019, USA
\\[\affilskip]
$^4$Max Planck Institut f{\"u}r Extraterrestrische Physik, Giessenbachstrasse 1, 85748 Garching, Germany
}

\pubyear{2017}
\volume{332}  
\jname{Astrochemistry VII - Through the Cosmos from Galaxies to Planets}
\editors{A.C. Editor, B.D. Editor \& C.E. Editor, eds.}
\begin{document}

\maketitle

\begin{abstract}
The youngest low-mass protostars are known to be chemically rich, accreting matter most vigorously, and producing the most powerful outflows. Molecules are unique tracers of these phenomena.
We use ALMA to study several outflow sources in the Serpens Main region. The most luminous source, Ser-SMM1, shows the richest chemical composition, but some complex molecules are also present in S68N. No emission from complex organics is detected toward Ser-emb 8N, which is the least luminous in the sample. We discuss whether these differences reflect an evolutionary effect or whether they are due to different physical structures.
We also analyze the outflow structure from these young protostars by comparing emission of CO and SiO. EHV molecular jets originating from SMM1-a,b and Ser-emb 8N contrast with no such activity from S68N, which on the other hand presents a complex outflow structure.
\end{abstract}

\firstsection 
\section{Introduction}
Deeply embedded Class 0 objects, the youngest low-mass protostars, present all kinds of chemical and kinematic complexity. The chemical richness is related to the physical processes associated with star formation within cold clouds of dust and gas. Deep inside, the gas is swept up and shocked by jets and outflows, while in parallel the material is heated, sublimates from the grains, and ultimately accretes onto the protostar through the disk. Class 0 sources accrete matter most vigorously, also producing the most powerful outflows \citep{Bontemps1996}, hence investigating this stage of protostellar evolution can provide a wealth of information about processes related to star formation.

Most of the outflowing molecular gas is present in relatively slow, massive and wide flows of entrained material. However, for some protostars an extremely high-velocity (EHV) component has been seen in e.g., CO, SiO, and SO, probably representing the original driving jet 
\citep[e.g.,][]{Bachiller1990, Santiago-Garcia2009, Hirano2010, Tafalla2010}.
Observing these very young jets could provide a tool to understand the launching mechanisms, interaction of the outflow with the envelope and chemical composition of the original jet.

Chemical complexity of the young protostars is particularly striking in the hot cores, where complex organic molecules can be detected. These molecules are of great importance in the context of understanding the origin of pre-biotic species, as many of those carbon-based molecules are building blocks of life as we know it. 
Since first detection of hot cores toward low-mass protostars \citep{vanDishoeck1995, Cazaux2003}, many more studies were dedicated to observe more sources that host complex organics \citep[e.g.,][]{Bergner2017} or targeting one of the protostars and scanning its submillimeter spectrum to detect a wealth of complex molecules \citep{Jorgensen2016}.

We use the Atacama Large Millimeter/submillimeter Array (ALMA) to study four Class 0 objects in the Serpens Main region \citep{Eiroa2008} located 436 pc away from the Sun \citep{Ortiz-Leon2017}. This small sample is attractive since it contains an intermediate mass protostar SMM1-a \citep[100 L\sol{};][]{Kristensen2012} as well as EHV molecular jets toward three protostars (SMM1-a, SMM1-b, and Ser-emb 8N). The high-resolution (0.5”) Band 6 (1.3 mm) ALMA observations yield spectra and maps showing various molecules. Here we present the analysis of CO and SiO emission in the outflows along with detections from complex organics molecules in the protostellar cores.

\section{Molecular jets}
\begin{figure}[h]
\begin{center}
\includegraphics[width=0.7\textwidth]{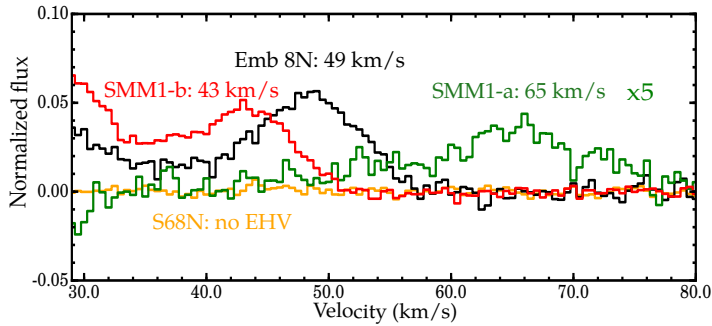} 
 \caption{Comparison of redshifted EHV components in the line wings of all observed outflow spectra with labels marking the velocity at which the EHV component reaches peak intensity. Flux density was normalized to the maximum CO intensity for each source. For SMM1-a the normalized flux was multiplied by factor of 5.}
   \label{fig6}
\end{center}
\end{figure}

Three outflows show a distinct high-velocity component in CO (2-1). Fig. \ref{fig6} presents a comparison of those components. The luminous source SMM1-a has the highest velocity for the peak wing intensity. However, as no luminosity estimate is known for the SMM1-b and Ser-emb 8N, we cannot conclude whether or not there is a relation of the bolometric luminosity and EHV velocity. Inclination can also significantly change the observed radial velocity. We further discuss each source separately.

\begin{figure}[h]
\begin{center}
\includegraphics[width=0.45\textwidth]{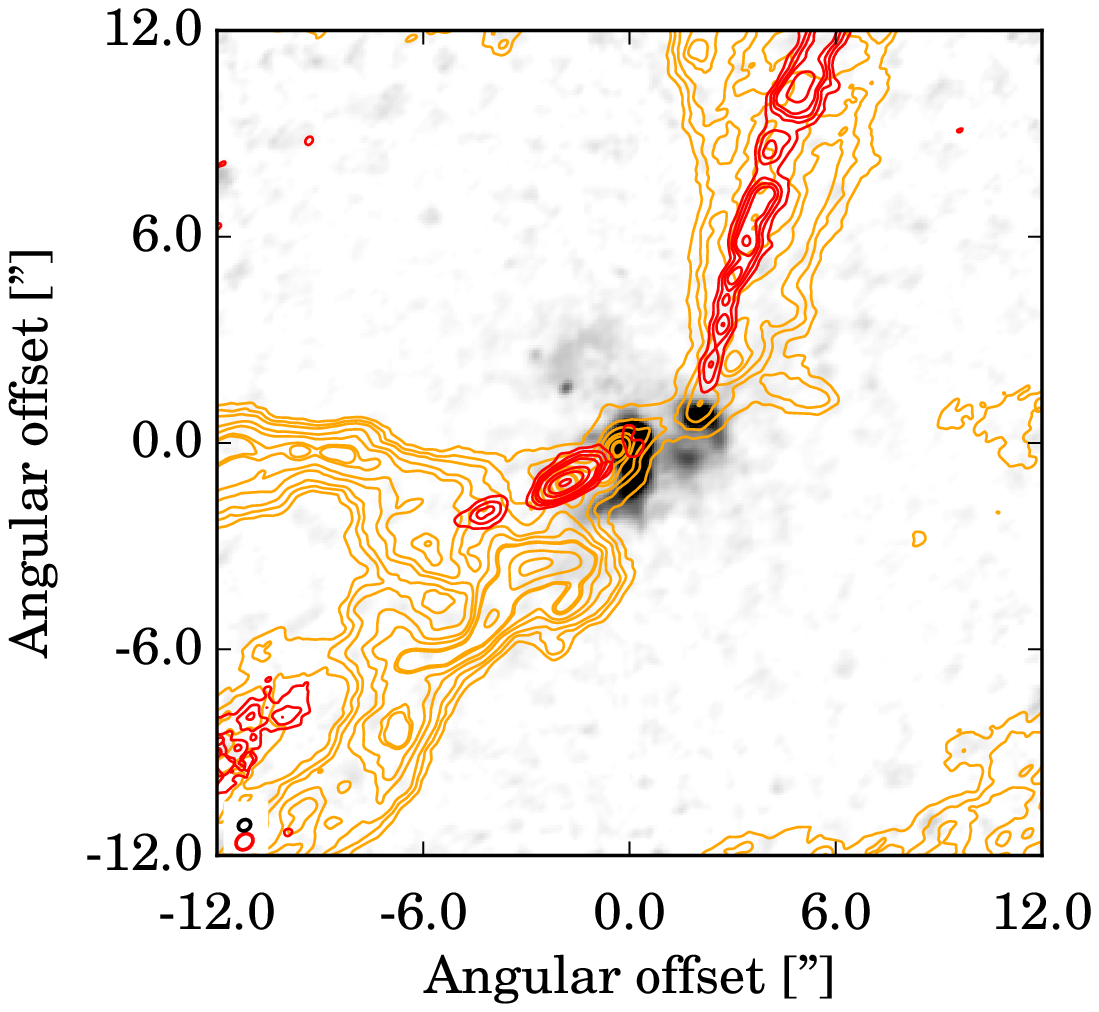} 
\includegraphics[width=0.45\textwidth]{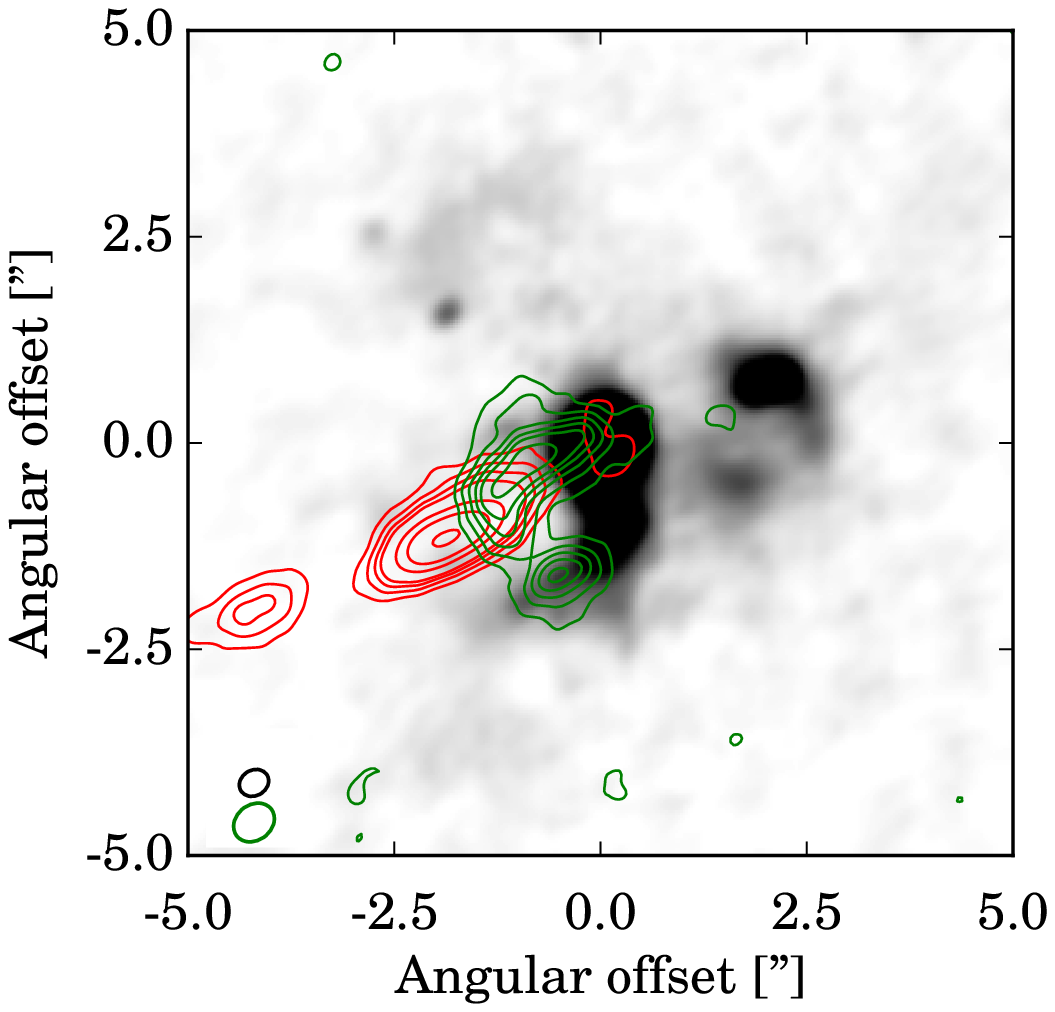}

 \caption{Serpens SMM1. Left: ALMA 1.3 mm continuum in grayscale with CO (2-1) integrated emission overlaid in contours. Red color represents EHV velocities from 44 to 70 km s$^{-1}$ and orange low-velocity outflows from 10 to 15 km s$^{-1}$. Right: Blow-up of central part with red contours showing CO (2-1) EHV emission from 60 to 70 km s$^{-1}$ and green contours show ${}^{13}$CS (5-4) integrated emission from 9 to 14 km s$^{-1}$. See also \cite{Hull2016,Hull2017}.}
   \label{fig1}
\end{center}
\end{figure}

Serpens SMM1-a is the most massive protostar in the Serpens molecular cloud. A CO (2-1) jet with velocities up to 90 km/s was observed by \cite{Hull2016}. The entrance of the outflow cavity wall is clearly delineated by the continuum dust emission, as well as in free-free emission \citep{Hull2016}. It appears that the molecular jet interacts heavily with the outflow cavity wall, as is indicated by enhanced ${}^{13}$CS emission on the side of the outflow where the EHV CO jet is currently pointing (see Fig. \ref{fig1}, right).

SMM1-b is another protostellar component in the SMM1 system, discovered by \cite{Choi2009}. \cite{Hull2016,Hull2017} present the high-velocity molecular jet toward the SMM1-b in CO and SiO respectively. The latter work also reveals that SMM1-b is a binary system itself with the eastern component powering the EHV jet. This source shows a clear example of the evolution of the fast collimated jet into a slow molecular wind in CO emission (see Fig \ref{fig1}, left).

Ser-emb 8 (S68N) is a protostar in the north-west part of the Serpens core. The outflow from S68N appears wide and evolved, with several spot shocks present in SiO, probably indicating interactions between the outflow and the surrounding cloud. This flow does not show signatures of EHV jets, consistent with the  more evolved nature of the central source.

\begin{figure}[h]
\begin{center}
\includegraphics[width=0.7\textwidth]{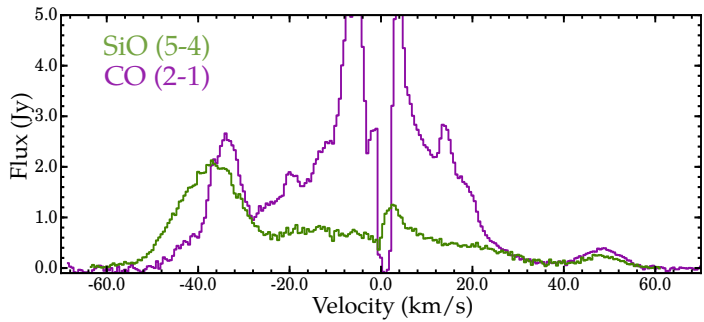}
 
 \caption{Integrated flux density over the total outflow area for Ser-emb 8N. SiO (5-4) is shown in green and CO (2-1) in purple.}
   \label{fig4}
\end{center}
\end{figure}

Protostar Ser-emb 8N is located about 10" north-east from S68N.
This source shows a high degree of collimation, even at low velocities, suggesting that the outflow is very young \citep{Arce2006}. There are symmetric high-velocity bullets on both sides of the jet, probably related to the outbursts of the accretion activity. Although geometrically symmetric, the blueshifted component appears as more powerful with stronger SiO and CO emission (Fig. \ref{fig4}).

\section{Complex organic molecules}
\begin{figure}[h]
\begin{center}
\includegraphics[width=0.4\textwidth]{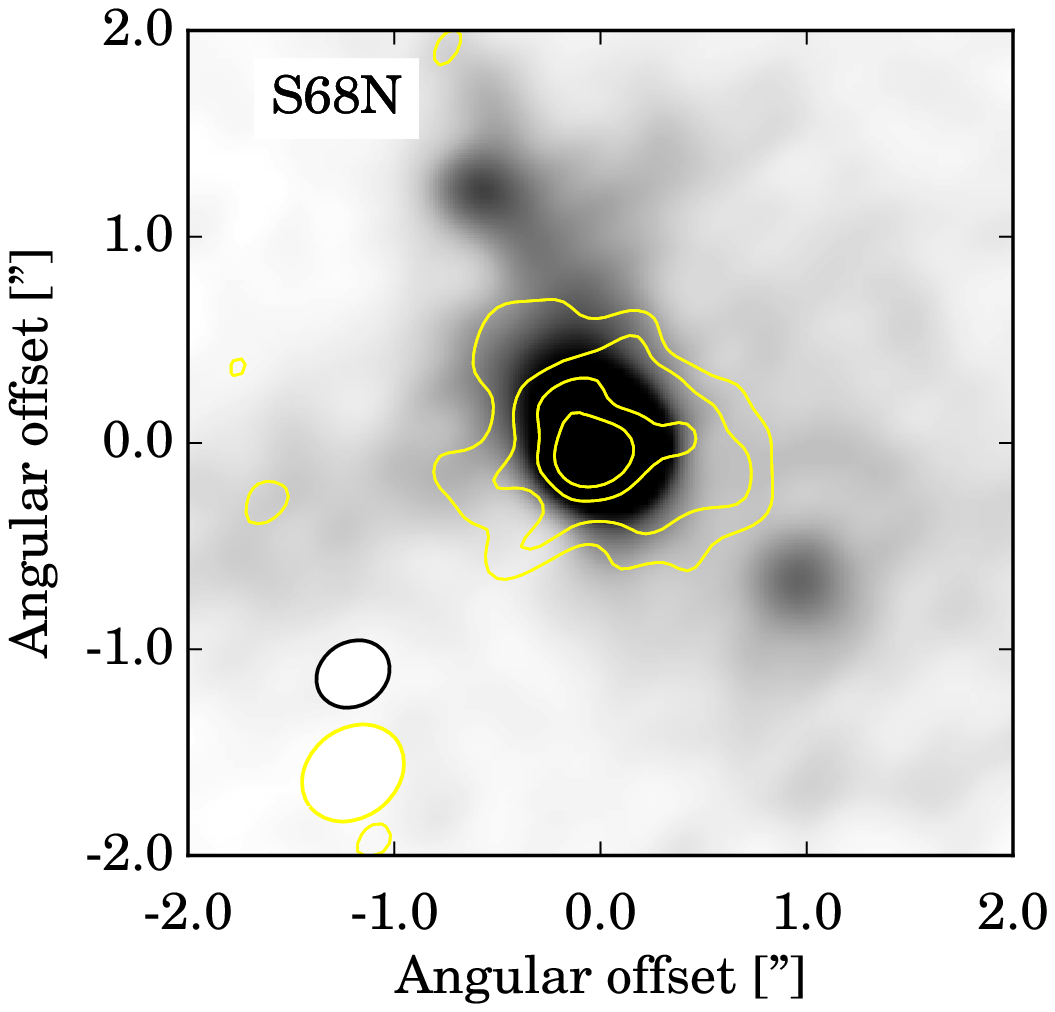} 
\includegraphics[width=0.4\textwidth]{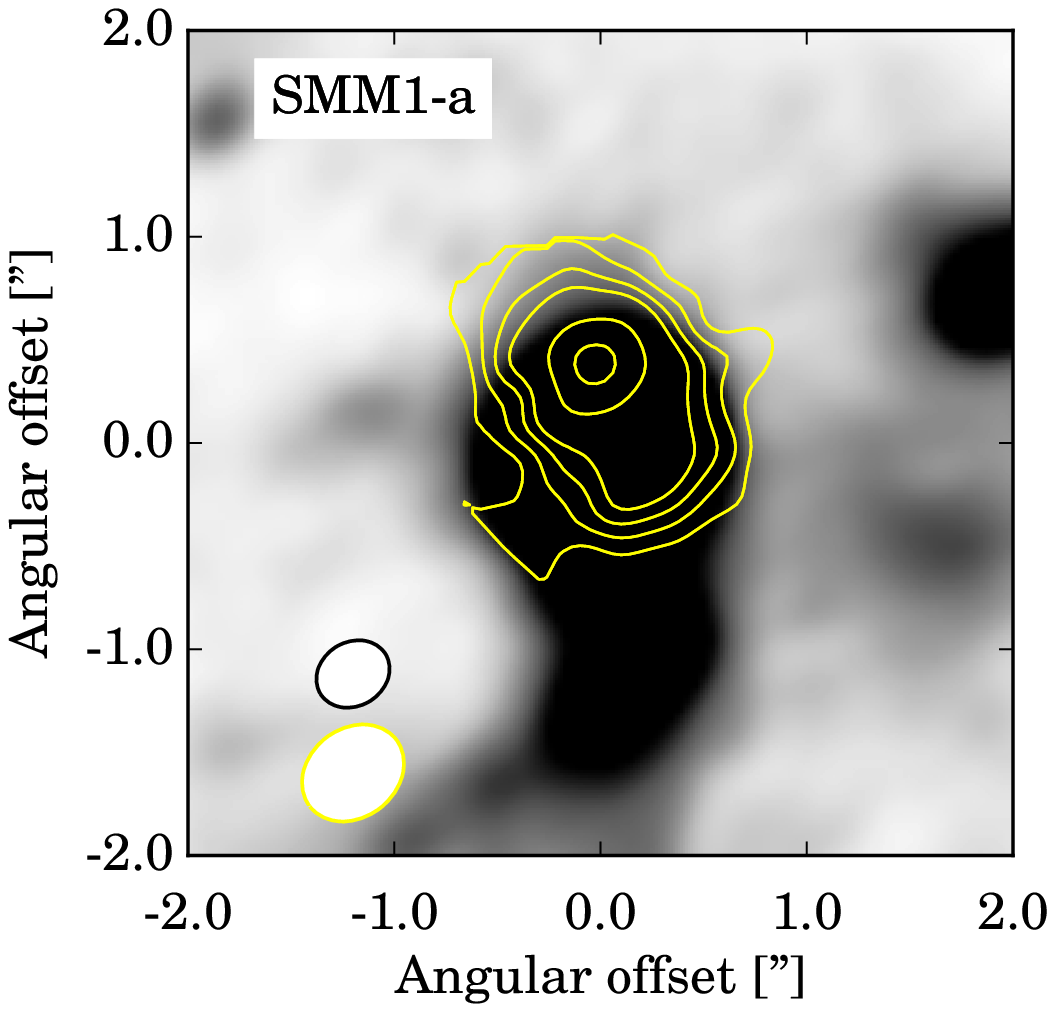}

 \caption{Left: 1.3 mm ALMA continuum emission in grayscale with an integrated emission of methyl formate in yellow contours toward S68N. Right: 1.3mm ALMA continuum emission in grayscale with an integrated emission of vinyl cyanide in yellow contours toward SMM1-a.}
   \label{fig5}
\end{center}
\end{figure}

With the high sensitivity of ALMA observations, a number of complex organic molecules are detected toward SMM1-a and S68N. Notably, no such emission is present in Ser-emb 8N and SMM1-b. Detection of carbon based complex molecules was reported for SMM1-a \citep{Oeberg2011}, while S68N is a new  detection of a hot corino, hinted by the presence of methanol in the source position by \cite{McMullin1994}.
Table \ref{tab1} shows the summary of detected molecules. Both SMM1-a and S68N have methanol, methyl formate, and dimethyl ether in their cores, while vinyl cyanide and ethylene glycol are present only toward SMM1-a. Emission from complex molecules is resolved, with peak of the emission located within the continuum source, with notable asymmetries for SMM1-a (Fig. \ref{fig5})
The luminosity of Ser-emb 8N and SMM1-b are poorly known, but based on their millimeter continuum they appear less massive than SMM1-a and S68N. Hence, the presence of the complex organic molecules may be dependent on the luminosity of the source, as the central protostar provides enough energy to heat up the dust in its immediate surroundings to the ice sublimation temperatures of $\sim$ 100 K out to a large enough distance.
It appears that, as we detect EHV emission toward SMM1-a alongside with complex organic molecules, detection of hot core chemistry as a signature of a more evolved source \citep{Santangelo2015, DeSimone2017} seems to fall.  However, the interpretation of the  CO jet from SMM1-a is not straightforward. There is no corresponding EHV emission in SiO, which is the case for Ser-emb 8N and SMM1-b, and the slow CO outflow has an opening angle of $\sim\ 90 \degree$, which can be interpreted as an indicator of a more evolved source.

\begin{table}
  \begin{center}
  \caption{Complex organic molecules detections}
  \label{tab1}
 {\scriptsize
  \begin{tabular}{|l|c|c|c|c|}\hline 
{\bf Name} & {\bf Formula} & {\bf SMM1-a} & {\bf S68N}
\\ \hline
Methanol& $\rm CH_2DOH$ & \cmark  & \cmark  \\
\hline
Methyl Formate& $\rm CH_3OCHO$ &  \cmark  &  \cmark 
\\ \hline
Dimethyl Ether &$\rm CH_3OCH_3$ &  \cmark   &  \cmark  
\\ \hline
Vinyl Cyanide &$\rm CH_2CHCN$ &  \cmark  &  \xmark 
\\ \hline
Ethylene Glycol & $\rm (CH_2OH)_2$ &  \cmark  &  \xmark \\ 
\hline

  \end{tabular}
  }
 \end{center}

\end{table}

\section{Summary}
We detect the high-velocity jets toward 3 out of 4 Class 0 objects, which suggests that these phenomena may be more common but detectable only at high sensitivity observations.  We also present detections and resolved images of complex organic molecules in the cores of two protostars. These results highlight new opportunities to study outflows from young protostars enabled by ALMA, as well as its unique place as the state-of-the-art astrochemistry tool, which brings us closer to understanding the path of complex molecules from interstellar clouds, through protostellar envelopes and disks, to planets.

\bibliography{my_bib.bib}

\begin{thebibliography}{}
\expandafter\ifx\csname natexlab\endcsname\relax\def\natexlab#1{#1}\fi

\bibitem[{Arce \& Sargent(2006)}]{Arce2006}
Arce, H.~G., \& Sargent, A.~I. 2006, Astrophysical Journal, 646, 1070

\bibitem[{Bachiller {et~al.}(1990)Bachiller, Martin-Pintado, Tafalla,
  Cernicharo, \& Lazareff}]{Bachiller1990}
Bachiller, R., Martin-Pintado, J., Tafalla, M., Cernicharo, J., \& Lazareff, B.
  1990, \aap, 231, 174

\bibitem[{{Bergner} {et~al.}(2017){Bergner}, {{\"O}berg}, {Garrod}, \&
  {Graninger}}]{Bergner2017}
{Bergner}, J.~B., {{\"O}berg}, K.~I., {Garrod}, R.~T., \& {Graninger}, D.~M.
  2017, \apj, 841, 120

\bibitem[{{Bontemps} {et~al.}(1996){Bontemps}, {Andre}, {Terebey}, \&
  {Cabrit}}]{Bontemps1996}
{Bontemps}, S., {Andre}, P., {Terebey}, S., \& {Cabrit}, S. 1996, \aap, 311,
  858

\bibitem[{Cazaux {et~al.}(2003)Cazaux, Tielens, Ceccarelli, Castets, Wakelam,
  Caux, Parise, \& Teyssier}]{Cazaux2003}
Cazaux, S., Tielens, A. G. G.~M., Ceccarelli, C., {et~al.} 2003, \apjl, 593,
  L51

\bibitem[{Choi(2009)}]{Choi2009}
Choi, M. 2009, \apj, 705, 1730

\bibitem[{{De Simone} {et~al.}(2017){De Simone}, {Codella}, {Testi},
  {Belloche}, {Maury}, {Anderl}, {Andr{\'e}}, {Maret}, \&
  {Podio}}]{DeSimone2017}
{De Simone}, M., {Codella}, C., {Testi}, L., {et~al.} 2017, \aap, 599, A121

\bibitem[{{Eiroa} {et~al.}(2008){Eiroa}, {Djupvik}, \& {Casali}}]{Eiroa2008}
{Eiroa}, C., {Djupvik}, A.~A., \& {Casali}, M.~M. 2008, {The Serpens Molecular
  Cloud}, ed. B.~{Reipurth}, 693

\bibitem[{Hirano {et~al.}(2010)Hirano, Ho, Liu, Shang, Lee, \&
  Bourke}]{Hirano2010}
Hirano, N., Ho, P. P.~T., Liu, S.-Y., {et~al.} 2010, \apj, 717, 58

\bibitem[{{Hull} {et~al.}(2016){Hull}, {Girart}, {Kristensen}, {Dunham},
  {Rodr{\'{\i}}guez-Kamenetzky}, {Carrasco-Gonz{\'a}lez}, {Cort{\'e}s}, {Li},
  \& {Plambeck}}]{Hull2016}
{Hull}, C.~L.~H., {Girart}, J.~M., {Kristensen}, L.~E., {et~al.} 2016, \apjl,
  823, L27

\bibitem[{{Hull} {et~al.}(2017){Hull}, {Girart}, {Tychoniec}, {Rao},
  {Cort{\'e}s}, {Pokhrel}, {Zhang}, {Houde}, {Dunham}, {Kristensen}, {Lai},
  {Li}, \& {Plambeck}}]{Hull2017}
{Hull}, C.~L.~H., {Girart}, J.~M., {Tychoniec}, {\L}., {et~al.} 2017, ArXiv
  e-prints, arXiv:1707.03827

\bibitem[{{J{\o}rgensen} {et~al.}(2016){J{\o}rgensen}, {van der Wiel},
  {Coutens}, {Lykke}, {M{\"u}ller}, {van Dishoeck}, {Calcutt}, {Bjerkeli},
  {Bourke}, {Drozdovskaya}, {Favre}, {Fayolle}, {Garrod}, {Jacobsen},
  {{\"O}berg}, {Persson}, \& {Wampfler}}]{Jorgensen2016}
{J{\o}rgensen}, J.~K., {van der Wiel}, M.~H.~D., {Coutens}, A., {et~al.} 2016,
  \aap, 595, A117

\bibitem[{{Kristensen} {et~al.}(2012){Kristensen}, {van Dishoeck}, {Bergin},
  {Visser}, {Y{\i}ld{\i}z}, {San Jose-Garcia}, {J{\o}rgensen}, {Herczeg},
  {Johnstone}, {Wampfler}, {Benz}, {Bruderer}, {Cabrit}, {Caselli}, {Doty},
  {Harsono}, {Herpin}, {Hogerheijde}, {Karska}, {van Kempen}, {Liseau},
  {Nisini}, {Tafalla}, {van der Tak}, \& {Wyrowski}}]{Kristensen2012}
{Kristensen}, L.~E., {van Dishoeck}, E.~F., {Bergin}, E.~A., {et~al.} 2012,
  \aap, 542, A8

\bibitem[{McMullin {et~al.}(1994)McMullin, Mundy, Wilking, Hezel, \&
  Blake}]{McMullin1994}
McMullin, J.~P., Mundy, L.~G., Wilking, B.~A., Hezel, T., \& Blake, G.~A. 1994,
  \apj, 424, 222

\bibitem[{{\"O}berg {et~al.}(2011){\"O}berg, van~der Marel, Kristensen, \& van
  Dishoeck}]{Oeberg2011}
{\"O}berg, K.~I., van~der Marel, N., Kristensen, L.~E., \& van Dishoeck, E.~F.
  2011, \apj, 740, 14

\bibitem[{Ortiz-Le{\'o}n {et~al.}(2017)Ortiz-Le{\'o}n, Loinard, Kounkel, Dzib,
  Mioduszewski, Rodr{\'{\i}}guez, Torres, Gonz{\'a}lez-L{\'o}pezlira, Pech,
  Rivera, Hartmann, Boden, Evans, Brice{\~n}o, Tobin, Galli, \&
  Gudehus}]{Ortiz-Leon2017}
Ortiz-Le{\'o}n, G.~N., Loinard, L., Kounkel, M.~A., {et~al.} 2017, \apj, 834,
  141

\bibitem[{{Santangelo} {et~al.}(2015){Santangelo}, {Codella}, {Cabrit},
  {Maury}, {Gueth}, {Maret}, {Lefloch}, {Belloche}, {Andr{\'e}}, {Hennebelle},
  {Anderl}, {Podio}, \& {Testi}}]{Santangelo2015}
{Santangelo}, G., {Codella}, C., {Cabrit}, S., {et~al.} 2015, \aap, 584, A126

\bibitem[{Santiago-Garc{\'{\i}}a {et~al.}(2009)Santiago-Garc{\'{\i}}a, Tafalla,
  Johnstone, \& Bachiller}]{Santiago-Garcia2009}
Santiago-Garc{\'{\i}}a, J., Tafalla, M., Johnstone, D., \& Bachiller, R. 2009,
  \aap, 495, 169

\bibitem[{Tafalla {et~al.}(2010)Tafalla, Santiago-Garc{\'{\i}}a, Hacar, \&
  Bachiller}]{Tafalla2010}
Tafalla, M., Santiago-Garc{\'{\i}}a, J., Hacar, A., \& Bachiller, R. 2010,
  \aap, 522, A91

\bibitem[{van Dishoeck {et~al.}(1995)van Dishoeck, Blake, Jansen, \&
  Groesbeck}]{vanDishoeck1995}
van Dishoeck, E.~F., Blake, G.~A., Jansen, D.~J., \& Groesbeck, T.~D. 1995,
  \apj, 447, 760

\end{thebibliography}

\end{document}